\documentclass[apjl]{emulateapj}
\usepackage{graphics}
\usepackage{apjfonts}
\usepackage{mathptmx}

\newcommand{\teff}{$T_{\!\mbox{\tiny\em eff}}$}
\newcommand{\yeff}{$y_{\!\mbox{\scriptsize eff}}$}

\newcommand{\hi}{H\,{\sc i}\rm}
\newcommand{\hii}{H\,{\sc ii}\rm}
\newcommand{\siii}{[S\,{\sc iii}]}
\newcommand{\nii}{[N\,{\sc ii}]}
\newcommand{\niii}{[N\,{\sc iii}]}
\newcommand{\oiii}{[O\,{\sc iii}]}
\newcommand{\oii}{[O\,{\sc ii}]}
\newcommand{\sii}{[S\,{\sc ii}]}
\newcommand{\ariii}{[Ar\,{\sc iii}]}
\newcommand{\neiii}{[Ne\,{\sc iii}]}

\newcommand{\te}{$T_e$}
\newcommand{\hbeta}{H$\beta$}

\newcommand{\lin}{$\,\lambda$}
\newcommand{\llin}{$\,\lambda\lambda$}
\slugcomment{}

\shorttitle{Metal-rich \hii\/ regions in M51}
\shortauthors{Bresolin, Garnett \& Kennicutt}

\begin{document}


\title{Abundances of metal-rich H\,II regions in
  M51$^1$}\footnotetext[1]{The data presented herein were obtained at
  the W.M. Keck Observatory, which is operated as a scientific
  partnership among the California Institute of Technology, the
  University of California and the National Aeronautics and Space
  Administration. The Observatory was made possible by the generous
  financial support of the W.M. Keck Foundation}

\author{Fabio Bresolin} \affil{Institute for Astronomy, 2680 Woodlawn
Drive, Honolulu, HI 96822; \it bresolin@ifa.hawaii.edu}

\author{Donald R. Garnett} \affil{Steward Observatory, University of
Arizona, 933 N. Cherry Ave., Tucson, AZ 85721;
\it dgarnett@as.arizona.edu}

\and 

\author{Robert C. Kennicutt, Jr.} \affil{Steward Observatory,
University of Arizona, 933 N. Cherry Ave., Tucson, AZ 85721;
\it robk@as.arizona.edu}

\begin{abstract}
We have obtained multi-object spectroscopy of \hii\/ regions in the
spiral galaxy M51 with the Keck~{\sc i} telescope and the Low
Resolution Imaging Spectrometer. For ten objects we have detected the
auroral line \nii\lin 5755, while \siii\lin 6312 has been measured in
seven of these. This has allowed us to measure the electron
temperature of the gas and to derive oxygen, sulfur and nitrogen
abundances for the ten \hii\/ regions. Contrary to the expectations
from previous photoionization models of a few \hii\/ regions in M51
and from strong-line abundance indicators, the O/H abundance is below
the solar value for most objects, with the most metal-rich \hii\/
regions, P\,203 and CCM~72, having log\,(O/H)\,=\,$-3.16$
[$\sim$$1.4\, {\rm (O/H)}_\odot$] and log\,(O/H)\,=\,$-3.29$
[$\sim$$1.0\, {\rm (O/H)}_\odot$], respectively.  The reduction of O/H
by factors up to two or three with respect to previous {\em indirect}
determinations has important consequences for the calibration of
empirical abundance indicators, such as $R_{23}$, in the abundance and
excitation range found in the central regions of spiral galaxies. The
abundance gradients in these galaxies can therefore be considerably
flatter than those determined by making use of such empirical
calibrations. The \hii\/ regions with a measured electron temperature
span the 0.19--1.04~$R_0$ range in galactocentric radius, and indicate
a shallow abundance gradient for M51: $-0.02 \pm 0.01$
dex\,kpc$^{-1}$.  The S/O abundance ratio is found to be similar to
previous determinations of its value in other spiral galaxies,
log\,(S/O)\,$\approx$\,$-1.6$.  Therefore, we find no evidence for a
variation in massive star initial mass function or nucleosynthesis at
high oxygen abundance. An overabundance of nitrogen is measured, with
log\,(N/O)\,$\simeq -0.6$.  Based on our new abundances we revise the
effective yield for M51, now found to be almost four times lower than
previous estimates, and we discuss this result within the context of
chemical evolution in galactic disks.  Features from Wolf-Rayet stars
(the blue bump at 4660~\AA\/ and the C\,{\sc iii} line at 5696~\AA)
are detected in a large number of \hii\/ regions in M51, with the
C\,{\sc iii}\lin 5696 line found preferentially in the central, most
metal-rich objects.
\end{abstract}



\section{Introduction}
Measuring the chemical abundances of gaseous nebulae is a crucial step
to understand the chemical evolution and the nucleosynthesis in spiral
galaxies. Historically, one of the main obstacles has been the
unavailability of electron temperatures ($T_e$) at high metallicity.
This is due to the enhanced cooling via far-IR lines causing the faint
auroral lines that are necessary to determine $T_e$ (e.g.~\oiii\lin
4363, \nii\lin 5755, \siii\lin 6312) to drop below detectability in
the spectra of \hii\/ regions. This difficulty affects the
observations of nebulae lying in the inner portions of most spiral
galaxies, where the oxygen abundance can exceed the solar value
[log\,(O/H)$_\odot$\,=\,$-3.31$, \citealt{allende01}]. In order to make this
important astrophysical problem tractable, strong-line methods have
been calibrated and almost universally adopted for the study of
abundance gradients in spirals and, more recently, to estimate
chemical abundances in high-redshift star-forming galaxies
(\citealt{alloin79}; \citealt{pagel79}; Edmunds \& Pagel~1984; Dopita
\& Evans~1986; \citealt{mcgaugh91}; D\'{\i}az \& P\'erez-Montero~2000;
\citealt{pilyugin00}; \citealt{kewley02}; \citealt{denicolo02};
\citealt{kobulnicky03}; Pettini \& Pagel~2004).  While these methods
can be calibrated empirically at low metallicity, nebular
photoionization models have provided the necessary calibration at high
metallicity (around the solar value and above), because of the lack of
reliable \hii\/ region abundances.

From the observations of a few extragalactic \hii\/ regions that were
believed to be extremely metal-rich (\citealt{kinkel94};
\citealt{diaz00b}; Castellanos, D\'{\i}az \& Terlevich~2002;
Kennicutt, Bresolin \& Garnett~2003) a disconcerting discrepancy
emerges between $T_e$-based oxygen abundances and those derived
through indirect means, the former being smaller by factors of a
few. These results suggest that the oxygen abundance in the inner
disks of spiral galaxies barely exceeds the solar value, while
strong-line methods would suggest abundances 2--3 times higher.

It is clear that this is exploration ground for large aperture
telescopes, given the faintess of the auroral lines one needs to
measure in order to obtain electron temperatures at high
metallicity. A first step in this direction has been taken by Garnett,
Kennicutt \& Bresolin~(2004), who used the 6.5m MMT telescope on
Mt.~Hopkins, Arizona, to obtain electron temperatures from the
\nii\lin 5755 line for two \hii\/ regions in M51, CCM~10 and
CCM~72. The \hii\/ regions in this galaxy have long been considered to
be extremely metal-rich, based on their very low excitation (weak
\oiii\lin\lin 4959,5007 lines relative to H$\beta$) and on the results
of photoionization models by \citet{diaz91}. The metallicity obtained
by \citet{garnett04}, however, is roughly a factor of three smaller,
barely exceeding the solar O/H value in the most metal-rich object,
CCM~72.

In this paper we present an enlarged sample of \hii\/ regions in M51,
observed with the Keck~{\sc i} telescope. The deeper spectra we
obtained allowed us to measure the \nii\lin 5755 line in ten \hii\/
regions, in conjunction with \siii\lin 6312 in seven of them.  The
larger number of objects and the measurement of two auroral lines
increases our confidence about the conclusions drawn in the previous
work. These observations, together with the data reduction, are
described in \S~2. The determination of electron temperatures and
chemical abundances is presented in \S~3. Major results on the oxygen,
nitrogen and sulfur abundances can be found in \S~4. The impact of the
resulting abundances on the calibration of strong-line methods is
discussed in \S~5. The Wolf-Rayet (WR) features detected among these
\hii\/ regions are briefly presented in \S~6. A few considerations on
effective yields and the gas fractions in M51 are addressed in \S~7,
and we summarize our conclusions in \S~8.

\section{Observations}

Spectroscopic observations of \hii\/ regions in M51 were carried out
on the night of April 24, 2003 at the W.~M.~Keck Observatory on Mauna
Kea, with the Keck~I telescope equipped with the Low Resolution
Imaging Spectrometer (\citealt{oke95}). The night was photometric,
with seeing at or below 1\arcsec.  Using the blue and red channels of
the spectrograph with a dichroic beam splitter, we obtained
simultaneous observations at wavelengths below and above
$\sim5300$~\AA.  For the blue setting a 600 lines mm$^{-1}$ grism
blazed at 4000~\AA\/ was used, providing spectra with a resolution of
4~\AA\/ FWHM on a mosaic of two 2K$\times$4K Marconi CCDs down to the
UV atmospheric cutoff.  For the red spectra the main observations,
covering roughly the 5300--7200~\AA\/ wavelength range, were carried
out with a 900 lines mm$^{-1}$ grating blazed at 5500~\AA\/ (3.5~\AA\/
resolution) and a 2K$\times$2K SITE CCD. Additional spectra with a 400
lines mm$^{-1}$ grating blazed at 8900~\AA\/ (7~\AA\/ resolution,
7000--10,000~\AA\/ approximate coverage) were obtained for the
measurement of the near-IR sulfur lines at
$\lambda\lambda$9069,9532~\AA.  However, due to the limited near-IR
extension of the calibration data, only the $\lambda$9069 fluxes are
useful, and the $\lambda$9532 fluxes were estimated from the
theoretical ratio $\lambda$9532/$\lambda$9069\,=\,2.44.

\begin{figure}[h]
\plotone{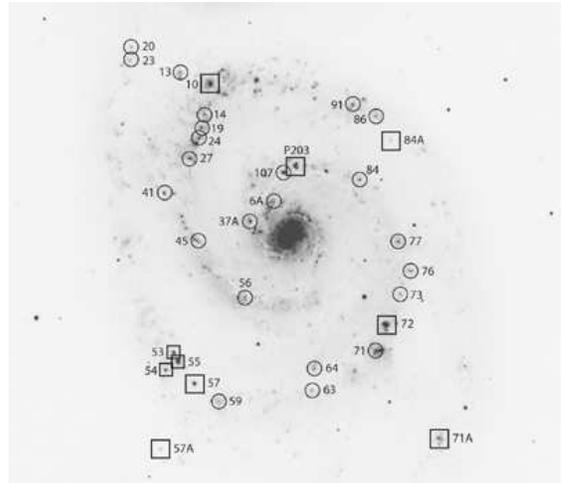}
\caption{Identification of the \hii\/ regions included in the Keck
  multi-object observations, plotted on top of a Kitt Peak 0.9m
  telescope H$\alpha$ image of M51 (image courtesy T. Rector; N at
  top, E at left). Numbers are from \citet{carranza69}, except for
  P\,203 (\citealt{petit96}). Squares mark the \hii\/ regions analyzed
  in this paper.  }
\label{map}
\end{figure}

Two slit masks were used for the multi-object spectroscopy, containing
21 and 16 1\farcs2-wide slits, respectively, with slit lengths
typically in the 10--20 arcsec range. The slits were oriented in the
North-South direction, and the airmass during the observations was
below 1.3.  Archival V, R and H$\alpha$ images of M51 provided the
necessary astrometry for the mask preparation. The target \hii\/
regions were selected among the brightest available in M51. A finding
chart is shown in Fig.~\ref{map}.  Five objects are in common with the
previous work by Bresolin, Kennicutt \& Garnett~(1999): CCM~10, 53,
55, 71A and 72 (catalog numbers from \citealt{carranza69}).

The observations analyzed in this paper consist of 2$\times$1800\,s
exposures in each of the 600/4000 and 900/5500 setups for the first
mask (objects east of the galactic nucleus), and 3$\times$1800\,s in
each setup for the second (west of the nucleus). Two 900\,s exposures
were used for the near-infrared (400/8900) spectra.  The spectrum from
each slit was treated as a separate spectrum, and reduced with common
{\sc iraf}\footnote{{\sc iraf} is distributed by the National Optical
Astronomy Observatories, which are operated by the Association of
Universities for Research in Astronomy, Inc., under cooperative
agreement with the National Science Foundation.} routines.  Standard
star spectra obtained at the start, middle and end of the night
provided the flux calibration. A standard extinction curve for Mauna
Kea was adopted for the atmospheric extinction correction
\citep{krisciunas87}.

The spectral region in common between the red and near-infrared
spectra was used for flux normalization, together with the flux in the
Paschen~9 and Paschen~10 lines relative to H$\beta$ when
available. This procedure introduced some additional uncertainty in
the \siii\lin\lin 9069,9532/\siii\lin 6312 line flux ratio, from which
the electron temperature in the intermediate-excitation zone is
derived (see \S 3). Given the insufficient useful overlap between the
blue and red spectra, emission line fluxes in the red spectra were
adjusted relative to the fluxes in the blue spectra by scaling the
flux in the H$\alpha$ line so that H$\alpha$/H$\beta$\,=\,3.00 as in
case B at \te\, = 6,000~K.  The ratio of the \nii\llin 6548,6583 to
\nii\lin 5755 line fluxes, from which the electron temperature in the
low-excitation zone is derived, is unaffected by scaling errors, as
these lines fall on the same spectra (the 900/5500 ones). Moreover,
this ratio is also largely insensitive to uncertainties in reddening
given the small wavelength baseline involved.

The reddening coefficient C(\hbeta) was determined from the Balmer
decrement using the H$\beta$, H$\gamma$ and H$\delta$ lines, adopting
the case B theoretical ratios at \te\, = 6,000~K, and the interstellar
reddening law of \citet{cardelli89}.  We have measured the \nii\lin
5755 auroral line in ten \hii\/ regions. Out of these objects,
\siii\lin 6312 has been measured in seven cases. The IR spectrum of
CCM~57A, however, fell outside of the area covered by the CCD, and
therefore for this object we could not measure T\siii\/ in the way
outlined in \S~3.  The \oiii\lin 4363 line remained undetected in all
objects, while \oii\lin 7325 was detected in CCM~10, CCM~57, CCM~71A
and CCM~72, however with a generally poor signal-to-noise ratio.
Besides the two nebulae in common with the study by \citet{garnett04},
this is the first time that auroral lines are measured in a large
number of \hii\/ regions in M51.  We will concentrate the rest of our
analysis on the ten objects with a measured \nii\lin 5755 line, as
information on electron temperatures and chemical abundances can be
obtained directly only for these \hii\/ regions.  The fluxes for their
most important lines, corrected for reddening and normalized to
F(\hbeta)\,=\,100, are presented in Table~1, where each \hii\/ region
is identified by its CCM number.  The \hii\/ region P\,203, 7\arcsec\/
West and 61\arcsec\/ North of the M51 nucleus, not included in the
Carranza et al.~(1969) catalog, is identified from \citet{petit96}.
The errors quoted in the Table reflect uncertainties in the flat
fielding and the flux calibration, as well as statistical errors,
which dominate in the case of the fainter lines (e.g. the auroral
lines).  The comparison concerning line fluxes for objects in common
with other authors, in particular \citet{bresolin99} and
\citet{garnett04} (CCM~10 and CCM~72), is generally good.  The
agreement with these works is typically at the 5 percent level or
better for the \nii\/ and \sii\/ lines, and at the 15 percent level or
better for the \oii, \oiii\/ and \siii\/ lines. A notable exception is
given by the \oiii\/ lines of CCM~53 (46 percent larger than in the
\citealt{bresolin99} paper), and the \siii\lin\lin 9069,9532 lines of
CCM~71A (a factor of two smaller). At least for the latter \hii\/
region, a well extended object with multiple emission peaks,
differences in pointing can be responsible for the disagreement.
Large differences are also found for CCM~10 when comparing with the
line intensities reported by \citet{diaz91}.

\begin{deluxetable*}{lcccccccccc}
\tabletypesize{\scriptsize}
\tablewidth{0pt}
\tablecolumns{11}
\tablenum{1}
\tablecaption{Dereddened Line Fluxes and Errors}
\tablehead{
\colhead{Line}            &
\colhead{CCM~10}    &
\colhead{CCM~53}    &
\colhead{CCM~54}    &
\colhead{CCM~55}    &
\colhead{CCM~57}    &
\colhead{CCM~57A}    &
\colhead{CCM~71A}    &
\colhead{CCM~72}    &
\colhead{CCM~84A}    &
\colhead{P\,203}    }
\startdata
\vspace{-2mm}
\\
[O\,{\sc ii}] \hfill 3727 & 126 $\pm$  9 & 129 $\pm$  9 & 115 $\pm$  8 &  78 $\pm$  5 & 114 $\pm$  8 & 104 $\pm$  7 & 147 $\pm$ 10 &  63 $\pm$  6 & 125 $\pm$ 12 &  32 $\pm$  2 \\
 
[Ne\,{\sc iii}] \hfill 3869 &  0.7 $\pm$ 0.1 &  0.6 $\pm$ 0.1 &  1.1 $\pm$ 0.2 &  0.8 $\pm$ 0.1 & \nodata & 12.8 $\pm$ 0.9 &  2.2 $\pm$ 0.2 &  0.4 $\pm$ 0.1 &  1.6 $\pm$ 0.4 & \nodata \\
 
H$\delta$ \hfill 4101 &  25 $\pm$  2 &  24 $\pm$  2 &  24 $\pm$  2 &  23 $\pm$  1 &  24 $\pm$  2 &  24 $\pm$  2 &  25 $\pm$  2 &  25 $\pm$  2 &  21 $\pm$  2 &  21 $\pm$  1 \\
 
H$\gamma$ \hfill 4340 &  48 $\pm$  3 &  49 $\pm$  3 &  46 $\pm$  3 &  46 $\pm$  3 &  47 $\pm$  3 &  48 $\pm$  3 &  46 $\pm$  3 &  47 $\pm$  3 &  45 $\pm$  3 &  43 $\pm$  3 \\
 
He\,{\sc i} \hfill 4471 &   2.4 $\pm$  0.2 &   2.9 $\pm$  0.2 &   2.4 $\pm$  0.3 &   2.1 $\pm$  0.1 &   2.9 $\pm$  0.2 &   5.6 $\pm$  0.4 &   2.7 $\pm$  0.2 &   2.1 $\pm$  0.2 &   4.8 $\pm$  0.5 & \nodata \\
 
[O\,{\sc iii}] \hfill 4959 &   4.4 $\pm$  0.3 &  11.8 $\pm$  0.7 &  14.2 $\pm$  0.9 &   6.7 $\pm$  0.4 &   5.7 $\pm$  0.3 &  60.0 $\pm$  3.5 &  16.1 $\pm$  1.0 &   2.4 $\pm$  0.1 &  29.8 $\pm$  1.8 &   1.7 $\pm$  0.2 \\
 
[O\,{\sc iii}] \hfill 5007 &  12.1 $\pm$  0.7 &  33.3 $\pm$  2.0 &  41.5 $\pm$  2.4 &  19.0 $\pm$  1.1 &  15.9 $\pm$  0.9 & 168.6 $\pm$  9.9 &  46.1 $\pm$  2.7 &   6.6 $\pm$  0.4 &  84.9 $\pm$  5.1 &   4.9 $\pm$  0.3 \\
 
[N\,{\sc ii}] \hfill 5755 & .50 $\pm$.04 & .54 $\pm$.07 & .65 $\pm$.07 & .43 $\pm$.04 & .68 $\pm$.09 & .48 $\pm$.08 & .79 $\pm$.07 & .28 $\pm$.04 & .92 $\pm$.15 & .15 $\pm$.02 \\
 
He\,{\sc i} \hfill 5876 &  7.7 $\pm$ 0.5 &  9.4 $\pm$ 0.6 & 10.0 $\pm$ 0.6 &  8.4 $\pm$ 0.5 & 10.1 $\pm$ 0.6 & 11.4 $\pm$ 0.7 &  9.3 $\pm$ 0.6 &  5.6 $\pm$ 0.4 & 13.1 $\pm$ 1.0 &  4.5 $\pm$ 0.3 \\
 
[S\,{\sc iii}] \hfill 6312 & .30 $\pm$.03 & .31 $\pm$.06 & .44 $\pm$.06 & .31 $\pm$.03 & .31 $\pm$.07 & .62 $\pm$.09 & \nodata & .16 $\pm$.03 & \nodata & \nodata \\
 
[N\,{\sc ii}] \hfill 6548 &  37 $\pm$  3 &  40 $\pm$  3 &  40 $\pm$  3 &  38 $\pm$  3 &  40 $\pm$  3 &  23 $\pm$  2 &  39 $\pm$  3 &  33 $\pm$  3 &  49 $\pm$  4 &  26 $\pm$  2 \\
 
H$\alpha$ \hfill 6563 & 300 $\pm$ 20 & 300 $\pm$ 20 & 300 $\pm$ 20 & 300 $\pm$ 20 & 300 $\pm$ 20 & 300 $\pm$ 20 & 300 $\pm$ 20 & 300 $\pm$ 27 & 300 $\pm$ 27 & 300 $\pm$ 20 \\
 
[N\,{\sc ii}] \hfill 6583 & 112 $\pm$  8 & 120 $\pm$  8 & 122 $\pm$  8 & 117 $\pm$  8 & 124 $\pm$  8 &  70 $\pm$  5 & 116 $\pm$  8 & 100 $\pm$  9 & 144 $\pm$ 13 &  79 $\pm$  5 \\
 
He\,{\sc i} \hfill 6678 &  2.0 $\pm$ 0.1 &  2.4 $\pm$ 0.2 &  2.6 $\pm$ 0.2 &  2.1 $\pm$ 0.1 &  2.3 $\pm$ 0.2 &  3.5 $\pm$ 0.3 &  2.7 $\pm$ 0.2 &  1.8 $\pm$ 0.2 &  3.1 $\pm$ 0.3 &  1.1 $\pm$ 0.1 \\
 
[S\,{\sc ii}] \hfill 6717 &  27 $\pm$  2 &  24 $\pm$  2 &  36 $\pm$  3 &  23 $\pm$  2 &  26 $\pm$  2 &  20 $\pm$  1 &  41 $\pm$  3 &  26 $\pm$  2 &  19 $\pm$  2 &  23 $\pm$  2 \\
 
[S\,{\sc ii}] \hfill 6731 &  19 $\pm$  1 &  17 $\pm$  1 &  27 $\pm$  2 &  18 $\pm$  1 &  19 $\pm$  1 &  14 $\pm$  1 &  28 $\pm$  2 &  20 $\pm$  2 &  15 $\pm$  1 &  18 $\pm$  1 \\
 
[Ar\,{\sc iii}] \hfill 7135 &  2.2 $\pm$ 0.2 &  2.7 $\pm$ 0.3 &  3.6 $\pm$ 0.4 &  2.6 $\pm$ 0.3 &  2.6 $\pm$ 0.3 & \nodata &  3.9 $\pm$ 0.5 &  1.4 $\pm$ 0.2 &  8.8 $\pm$ 2.8 &  1.7 $\pm$ 0.2 \\
 
[S\,{\sc iii}] \hfill 9069 &  17 $\pm$  3 &  18 $\pm$  4 &  17 $\pm$  3 &  21 $\pm$  4 &  14 $\pm$  3 & \nodata &  13 $\pm$  3 &  14 $\pm$  3 &  17 $\pm$  3 &   9 $\pm$  2 \\
 
C(H$\beta$) &       0.25 $\pm$0.05 & 0.10 $\pm$0.05 & 0.15 $\pm$0.05 & 0.05 $\pm$0.05 & 0.15 $\pm$0.05 & 0.05 $\pm$0.05 & 0.25 $\pm$0.05 & 0.35 $\pm$0.10 & 0.60 $\pm$0.10 & 0.21 $\pm$0.05 \\
 
EW(H$\beta$)  (\AA)    &    51 &    92 &   144 &    74 &   155 &   300 &    96 &    61 &   203 &    47 \\

\enddata
\end{deluxetable*}


\section{Electron temperatures and abundances}

As in our previous work on M101 (\citealt{kennicutt03}) and on the M51
\hii\/ regions CCM~72 and CCM~10 (\citealt{garnett04}), we adopt a
three-zone model to describe the ionization structure of an \hii\/
region. These zones are characterized by different electron
temperatures \te, each referring to different, coexisting atomic
ionization stages: \oii, \nii, \sii\/ in the low-ionization zone,
\siii, \ariii\/ in the intermediate-ionization zone, and \oiii,
\neiii\/ in the high-ionization zone. The photoionization models of
\citet{garnett92} predict simple scaling relations between the
temperature in the different zones, applicable in a wide T\oiii\/
range (2,000--18,000~K):

\begin{center}
\begin{equation}
{\rm T[S\,III] = 0.83\, T[O\,III]\,+\,1700~{\rm K}},
\end{equation}
\end{center}

\begin{center}
\begin{equation}
{\rm T[N\,II] = T[O\,II] = 0.70\, T[O\,III]\,+\,3000~{\rm K}}.
\end{equation}
\end{center}

\noindent
Observational evidence in support of these relationships, in
particular Eq.~(1), has been provided by a few authors (e.g.\/
\citealt{garnett97}; \citealt{kennicutt03}). One could consider
alternative formulations, for example the relationship between T\oii\/
and T\oiii\/ given by \citet{izotov94}, based on the grid of
photoionization models by \citet{stasinska90}. For the same measured
value of T\oii, their Eq. (4) would lead to T\oiii\/ lower by a few
hundred degrees relative to the \citet{garnett92} equation. However,
if we limit the Stasinska models to \teff\/ $<$ 40,000~K for the
ionizing stars, as seems appropriate for the \hii\/ regions in M51, we
obtain good agreement betwen the two different sets of photoionization
models.

We established from the \sii\lin 6717/\sii\lin 6731 line ratio that
all the \hii\/ regions are in the low-density regime ($N_e < 150$
cm$^{-3}$). Subsequently, using the five-level atom program {\em
nebular} implemented in {\sc iraf/stsdas} (\citealt{shaw95}), we
determined the electron temperatures T\nii\/ (ten objects) and
T\siii\/ (six objects) from the emission line ratios \nii\llin
6548,6583\,/\,\nii\lin 5755 and \siii\lin\lin 9069,9532\,/\,\siii\lin
6312, respectively. The results are summarized in Table~2. As in the
M101 \citet{kennicutt03} paper, we have updated the S\,{\sc iii}
collisional strengths used by {\em nebular} adopting the results by
\cite{tayal99}.

\begin{figure}[h]
\plotone{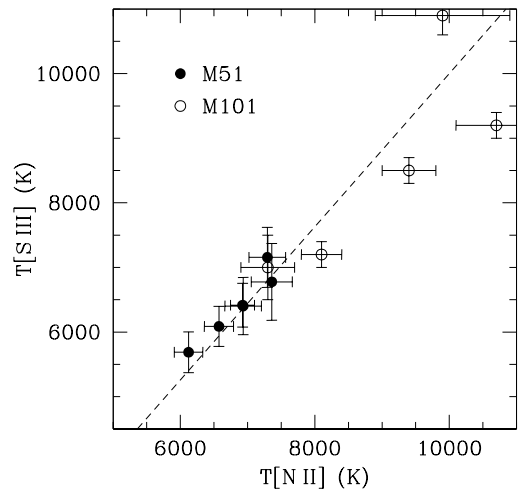}
\caption{Relationship between T\nii\/ and T\siii\/ measured for \hii\/
  regions in M51 (this work: full dots) and M101
  (\citealt{kennicutt03}: open dots). The model prediction from
  \citet{garnett92} is drawn as a dashed line.}
\label{temp}
\end{figure}

\begin{deluxetable}{lcc}
\tabletypesize{\scriptsize}
\tablecolumns{3}
\tablewidth{0pt}
\tablenum{2}
\tablecaption{Measured Electron Temperatures (K)}

\tablehead{
\colhead{ID}            &
\colhead{T[N\,{\sc ii}]}	&
\colhead{T[S\,{\sc iii}]}	\\
\colhead{}            &
\colhead{(K)}	&
\colhead{(K)}}

\startdata
\vspace{-2mm}
\\
CCM~10  &  6900 $\pm$ 200  &  6400 $\pm$ 300 \\
CCM~53  &  6900 $\pm$ 300  &  6400 $\pm$ 400 \\
CCM~54  &  7300 $\pm$ 300  &  7200 $\pm$ 500 \\
CCM~55  &  6600 $\pm$ 200  &  6100 $\pm$ 300 \\
CCM~57  &  7400 $\pm$ 300  &  6800 $\pm$ 600 \\
CCM~57A  & 7900 $\pm$ 400  &  \nodata \\
CCM~71A  & 7700 $\pm$ 200  &  \nodata \\
CCM~72  &  6100 $\pm$ 200  &  5700 $\pm$ 300 \\
CCM~84A  & 7700 $\pm$ 400  &  \nodata \\
P\,203   &  5600 $\pm$ 200  &  \nodata \\
\enddata
\end{deluxetable}

A good agreement is found with the independent T\nii\/ determinations
by \citet{garnett04} for CCM~10 [7400 (+1000,-600) K] and for CCM~72
(6000 $\pm$ 300 K). To verify the \hii\/ region zone model adopted for
the following analysis, we have considered the correlation between
T\nii\/ and T\siii\/ for the six M51 objects where both temperatures
are available. As Fig.~\ref{temp} shows, the observed relation agrees
very well with the expectations based on the \citet{garnett92} models
(dashed line). This result is encouraging, since it lends support to
the \hii\/ region zone model adopted here, indicating that the
electron temperatures in the different excitation zones can be well
established even when only one auroral line (e.g. \nii\lin 5755 at low
excitation) is available.  We have therefore calculated the 'missing'
temperature for the high-ionization zone T\oiii, using both equations
(1) and (2), and taking a weighted average for the result when both
T\nii\/ and T\siii\/ are available. This gives us electron temperature
estimates for the high-, intermediate- and low-ionization zones as
described above; the results are listed in Table~3. For six of the ten
objects listed we have two independent estimates of T\oiii, which we
have averaged, from our measured T\nii\/ and T\siii. We note that the
2$\sigma$ value of T\siii\/ $\approx$ 5400~K for CCM~72 from
\citet{garnett04} is very close to our result (5700 $\pm$ 300~K). The
lowest temperature in our \hii\/ region sample is measured for P~203,
close to the galactic center, which, at T\nii\,=\,5600~K, is the
coolest extragalactic nebula to date where a direct electron
temperature exists.

\begin{deluxetable}{lccc}
\tabletypesize{\scriptsize}
\tablecolumns{4}
\tablewidth{0pt}
\tablenum{3}
\tablecaption{Adopted Electron Temperatures (K)}
\tablehead{
\colhead{ID}            &
\colhead{T(O$^+$, N$^+$, S$^+$)}	&
\colhead{T(S$^{+2}$, Ar$^{+2}$)}	&
\colhead{T(O$^{+2}$, Ne$^{+2}$)}	\\
\colhead{}            &
\colhead{(K)}	&
\colhead{(K)}	&
\colhead{(K)}}
\startdata
\vspace{-2mm}
\\
CCM~10  &  6900 $\pm$ 200  &  6400 $\pm$ 200 & 5600 $\pm$ 200 \\
CCM~53  &  6900 $\pm$ 300  &  6400 $\pm$ 300 & 5600 $\pm$ 300 \\
CCM~54  &  7300 $\pm$ 300  &  7200 $\pm$ 300 & 6300 $\pm$ 300 \\
CCM~55  &  6600 $\pm$ 200  &  6100 $\pm$ 200 & 5200 $\pm$ 200 \\
CCM~57  &  7400 $\pm$ 300  &  6800 $\pm$ 300 & 6200 $\pm$ 400 \\
CCM~57A  & 7900 $\pm$ 400  &  7500 $\pm$ 400 & 7000 $\pm$ 600 \\
CCM~71A  & 7700 $\pm$ 200  &  7300 $\pm$ 200 & 6800 $\pm$ 400 \\
CCM~72  &  6100 $\pm$ 200  &  5700 $\pm$ 200 & 4600 $\pm$ 200 \\
CCM~84A  & 7700 $\pm$ 400  &  7300 $\pm$ 400 & 6900 $\pm$ 600 \\
P~203  &   5600 $\pm$ 200  &  4800 $\pm$ 200 & 3700 $\pm$ 300 \\
\enddata
\end{deluxetable}

The ionic and element abundances derived from the electron
temperatures in Table~3 using the five-level atom program are listed
in Table~4 and 5, respectively. The associated errors derive mostly
from the uncertainty in the electron temperatures. The uncertainty in
the sulfur abundance is largest for CCM~57A, where we had to rely on
\lin 6312 alone for the S$^{+2}$ ionic abundance, due to the lack of
the IR spectrum mentioned in \S~2.

\begin{figure}[h]
\plotone{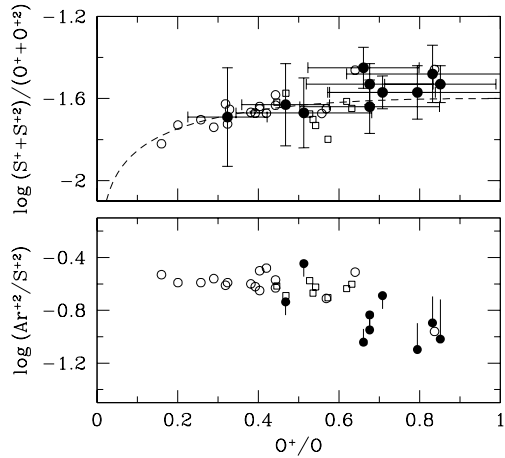}
\caption{{\em Top:} The ion ratio
(S$^+$\,+\,S$^{+2}$)\,/\,(O$^+$\,+\,O$^{+2}$) plotted against the
oxygen fractional ionization O$^+$/O for our sample of objects in M51
(full dots with error bars), and for objects studied in M101
(\citealt{kennicutt03}, open circles) and in NGC~2403
(\citealt{garnett97}, open squares). Equation (3) for the sulfur
ionization correction factor provides the dashed curve.  {\em Bottom:}
Ar$^{+2}$/S$^{+2}$ plotted against O$^+$/O for the same sample of
objects as above. A correlation with ionization seems to occur at
large O$^+$/O values. The vertical bars show the estimated correction
factors for the M51 \hii\/ regions (see text for explanation).  }
\label{ion_s}
\end{figure}

Ionization correction factors for sulfur are expected to be small in
low-ionization nebulae, as is the case here. This is verified in the
top panel of Fig.~\ref{ion_s}, where the ion ratio
(S$^+$\,+\,S$^{+2}$)/(O$^+$\,+\,O$^{+2}$) is plotted against the
oxygen ionization fraction O$^+$/O. The M51 data are in general
agreement with the trend observed for \hii\/ regions in M101 and
NGC~2403 (open circles and squares, respectively), extended to higher
O$^+$/O, even if a couple of objects (CCM~55 and 57) display a slight
offset from the average relation.  The dashed line in Fig.~\ref{ion_s}
reproduces the sulfur ionization correction formula
(\citealt{stasinska78}; \citealt{french81}):

\begin{center}
\begin{equation}
\frac{{\rm S}^+ + {\rm S}^{+2}}{{\rm S}} = \left[ 1 - \left(
  1-\frac{{\rm O}^+}{{\rm O}} \right)^\alpha \right] ^{1/\alpha},
\end{equation}
\end{center}

\noindent
where we have used $\alpha=2.5$ and log\,(S/O)$=-1.6$, which are the
 parameters we adopted in our M101 paper. Our new observations in M51
 do not justify a change in the parameters that appear in the
 ionization correction formula.

\begin{figure}[h]
\plotone{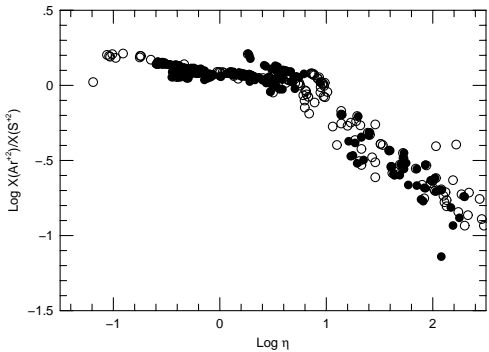}
\caption{The predicted correlation between the
X(Ar$^{+2}$)/X(S$^{+2}$) ionic fraction and the radiation softness
parameter $\eta$ from the models of Stasinska et al.~(2001). The
latter are plotted for an upper limit of the initial mass function of
120~M$_\odot$ ({\em filled circles}) and 30~M$_\odot$ ({\em open
circles}), for both extended and instantaneous bursts. The ionization
correction factor used to obtain the abundance ratio Ar/S from
Ar$^{+2}$/S$^{+2}$ can be estimated from the reciprocal of
X(Ar$^{+2}$)/X(S$^{+2}$) in this plot. }
\label{icf}
\end{figure}

In the case of argon the situation appears more complex. Unlike
sulfur, only one ionization stage (Ar$^{+2}$) is observed for this
atomic species in the optical spectra of \hii\/ regions, and in
low-ionization nebulae the amount of unseen Ar$^+$ can become
non-negligible. In our previous work on M101, we established that
Ar$^{+2}$\,/\,S$^{+2}$ is a good tracer of Ar\,/\,S, based on the near
independence of this ratio on the oxygen ionization fraction (at the
same time we showed that Ar$^{+2}$\,/\,O$^{+2}$ is a bad tracer of
Ar\,/\,O, and that neon has similar problems). The bottom plot of
Fig.~\ref{ion_s} suggests, however, that at high O$^+$\,/\,O (low
ionization) Ar$^{+2}$\,/\,S$^{+2}$ deviates from the roughly constant
value found at higher ionization. This is likely an effect of the
increase of Ar$^+$ with respect to Ar$^{+2}$ at low excitation.  There
is, in fact, a good correlation between Ar$^{+2}$\,/\,S$^{+2}$ and the
hardness of the radiation field, as measured by the parameter $\eta =
({\rm O}^+ / {\rm O}^{+2}) / ({\rm S}^+ / {\rm S}^{+2})$
(\citealt{vilchez88}; see Fig.~\ref{icf}). The ionization correction
factor necessary to estimate Ar\,/\,S from Ar$^{+2}$\,/\,S$^{+2}$ was
obtained from the corresponding quantities in the models of
\citet{stasinska01}, and indicate that at low excitation the
correction can amount to $+0.2$ to $+0.3$ dex.  The shift for each
\hii\/ region in the diagram thus determined is shown by the vertical
bars in Fig.~\ref{ion_s}. This ionization correction scheme also shows
that at high excitation (as for most of the M101 and NGC~2403 \hii\/
regions plotted) a moderate negative correction ($\simeq -0.1$ dex)
should be applied to the observed Ar$^{+2}$\,/\,S$^{+2}$ ratios in
order to recover the real Ar\,/\,S ratios. This will remove most of
the apparent decreasing trend of Ar\,/\,S with oxygen ionization
fraction suggested at first from Fig.~\ref{ion_s}.

\begin{deluxetable*}{lccccccc}
\tabletypesize{\scriptsize}
\tablecolumns{8}
\tablewidth{0pt}
\tablenum{4}
\tablecaption{Ionic Abundances}
\tablehead{
\colhead{ID}            &
\colhead{O$^+$/H$^+$}	&
\colhead{O$^{+2}$/H$^+$}	&
\colhead{N$^+$/H$^+$}	&
\colhead{S$^+$/H$^+$}	&
\colhead{S$^{+2}$/H$^+$}	&
\colhead{Ne$^{+2}$/H$^+$}	&
\colhead{Ar$^{+2}$/H$^+$}	}
\startdata
\vspace{-2mm}
\\
CCM~10  & (3.1$\pm$0.6)$\times 10^{-4}$ & (5.5$\pm$1.4)$\times 10^{-5}$ & (6.8$\pm$0.7)$\times 10^{-5}$ & (3.0$\pm$0.3)$\times 10^{-6}$ & (7.5$\pm$1.3)$\times 10^{-6}$ & (1.6$\pm$0.5)$\times 10^{-5}$ & (7.2$\pm$1.5)$\times 10^{-7}$ \\
CCM~53  & (3.1$\pm$0.8)$\times 10^{-4}$ & (1.5$\pm$0.6)$\times 10^{-4}$ & (7.3$\pm$1.2)$\times 10^{-5}$ & (2.7$\pm$0.4)$\times 10^{-6}$ & (7.9$\pm$1.9)$\times 10^{-6}$ & (1.5$\pm$0.7)$\times 10^{-5}$ & (8.9$\pm$2.6)$\times 10^{-7}$ \\
CCM~54  & (2.1$\pm$0.5)$\times 10^{-4}$ & (1.0$\pm$0.3)$\times 10^{-4}$ & (6.2$\pm$0.9)$\times 10^{-5}$ & (3.5$\pm$0.5)$\times 10^{-6}$ & (5.6$\pm$1.1)$\times 10^{-6}$ & (1.2$\pm$0.5)$\times 10^{-5}$ & (8.2$\pm$2.0)$\times 10^{-7}$ \\
CCM~55  & (2.6$\pm$0.6)$\times 10^{-4}$ & (1.4$\pm$0.5)$\times 10^{-4}$ & (8.5$\pm$1.2)$\times 10^{-5}$ & (3.3$\pm$0.4)$\times 10^{-6}$ & (1.1$\pm$0.2)$\times 10^{-5}$ & (3.2$\pm$1.4)$\times 10^{-5}$ & (1.0$\pm$0.2)$\times 10^{-6}$ \\
CCM~57  & (2.1$\pm$0.6)$\times 10^{-4}$ & (4.4$\pm$1.8)$\times 10^{-5}$ & (6.1$\pm$1.0)$\times 10^{-5}$ & (2.4$\pm$0.4)$\times 10^{-6}$ & (5.5$\pm$1.7)$\times 10^{-6}$ & \nodata & (7.0$\pm$2.6)$\times 10^{-7}$ \\
CCM~57A  & (1.2$\pm$0.4)$\times 10^{-4}$ & (2.6$\pm$1.4)$\times 10^{-4}$ & (2.7$\pm$0.6)$\times 10^{-5}$ & (1.5$\pm$0.3)$\times 10^{-6}$ & (6.2$\pm$3.0)$\times 10^{-6}$ & (8.2$\pm$5.6)$\times 10^{-5}$ & \nodata \\
CCM~71A  & (2.0$\pm$0.4)$\times 10^{-4}$ & (8.1$\pm$2.4)$\times 10^{-5}$ & (4.8$\pm$0.6)$\times 10^{-5}$ & (3.3$\pm$0.3)$\times 10^{-6}$ & (4.1$\pm$0.5)$\times 10^{-6}$ & (1.7$\pm$0.6)$\times 10^{-5}$ & (8.4$\pm$1.1)$\times 10^{-7}$ \\
CCM~72  & (4.1$\pm$1.3)$\times 10^{-4}$ & (1.0$\pm$0.4)$\times 10^{-4}$ & (9.6$\pm$1.5)$\times 10^{-5}$ & (4.8$\pm$0.7)$\times 10^{-6}$ & (9.0$\pm$1.9)$\times 10^{-6}$ & (3.8$\pm$2.2)$\times 10^{-5}$ & (7.2$\pm$1.8)$\times 10^{-7}$ \\
CCM~84A  & (1.7$\pm$0.6)$\times 10^{-4}$ & (1.6$\pm$1.0)$\times 10^{-4}$ & (6.2$\pm$1.3)$\times 10^{-5}$ & (1.6$\pm$0.3)$\times 10^{-6}$ & (5.3$\pm$1.1)$\times 10^{-6}$ & (1.3$\pm$1.0)$\times 10^{-5}$ & (1.9$\pm$0.5)$\times 10^{-6}$ \\
P\,203  & (3.2$\pm$0.8)$\times 10^{-4}$ & (3.6$\pm$2.9)$\times10^{-4}$ & (1.1$\pm$0.2)$\times 10^{-4}$ & (6.0$\pm$0.8)$\times 10^{-6}$ & (9.8$\pm$1.7)$\times 10^{-6}$ & \nodata & (1.8$\pm$0.4)$\times 10^{-6}$  \\
\enddata
\end{deluxetable*}


\section{Results for oxygen, sulfur and nitrogen}

The somewhat surprising result we obtain concerns the typical oxygen
abundance of the M51 \hii\/ regions studied here: for most objects O/H
is below the solar value [log\,(O/H)$_\odot$\,=\,$-3.31$, Allende
Prieto et al.~2001], and reaches $\sim$1.4 times the solar value for
P\,203 [log\,(O/H)\,=\,$-3.16$], the most oxygen-rich \hii\/ region in
our sample.  This contrasts with the results of detailed
photoionization models by \citet{diaz91}, who estimated highly
super-solar abundances, between log\,(O/H)\,=\,$-2.9$ and
log\,(O/H)\,=\,$-2.6$ for their objects (CCM~10 and CCM~72 are in
common with our sample), and thus confirms the findings of
\citet{garnett04}. The concept of M51 as a very metal-rich (largely
over-solar) spiral galaxy, supported since the early 1990's, based on
the above-mentioned models and the accepted calibrations of several
empirical abundance determination methods, in particular $R_{23}$
(\citealt{vila92}), is not fully supported by our data. The results
for P\,203, on the other hand, suggest that the central region of the
galaxy, where the excitation is very low, contains nebulae with a O/H
ratio at least 40\% larger then the solar value. The discussion of
additional low-excitation \hii\/ regions near the M51 nucleus is
deferred to the end of \S~5.

\begin{figure}[h]
\plotone{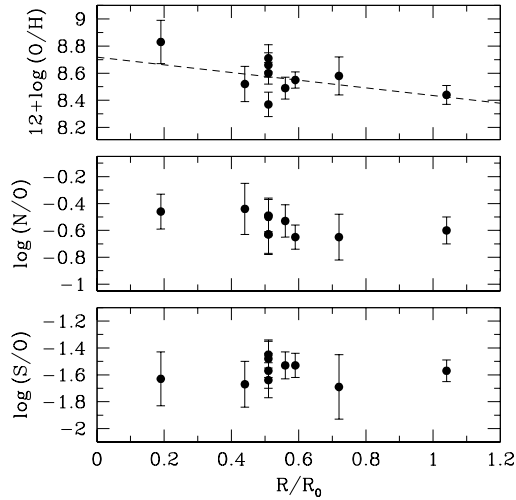}
\caption{Radial trends of 12\,+\,log\,(O/H) ({\em top}), (b) log\,(N/O)
  ({\em middle}) and log\,(S/O) ({\em bottom}). The dashed line
  represents the linear fit to the O/H radial gradient (equation~4).}
\label{radial}
\end{figure}

Column 2 in Table~5 lists the deprojected distance from the galaxy
center in units of the isophotal radius ($R_0$ = 5.4 arcmin,
corresponding to 13.2 kpc if the distance of 8.4~Mpc is adopted from
\citealt{feldmeier97}). Our data span the disk from $R/R_0=0.19$ to
$R/R_0=1.04$, but with the majority of the objects concentrated around
$R/R_0=0.5$.  As a result the radial trends of abundances and
abundance ratios depend heavily on the \hii\/ regions at the two
opposite extremes of the range, P\,203 (inner) and CCM~71A (outer).
We plot the radial distribution of O/H, N/O and S/O in
Fig.~\ref{radial}, which indicates a rather shallow O/H gradient in
M51. The dashed line represents the linear regression to the data
points:

\begin{center}
\begin{equation}
{\rm 12 + \log (O/H) = 8.72\; (\pm 0.09) -0.28\; (\pm 0.14)\;} R/R_0
\end{equation}
\end{center}

\noindent

\begin{deluxetable}{lcccc}
\tabletypesize{\scriptsize}
\tablecolumns{5}
\tablewidth{0pt}
\tablenum{5}
\tablecaption{Total abundances}
\tablehead{
\colhead{ID}            &
\colhead{$R/R_0$}       &
\colhead{12+log\,(O/H)}	&
\colhead{log\,(N/O)}	&
\colhead{log\,(S/O)}	}
\startdata
\vspace{-2mm}
\\
CCM~10  & 0.59 & 8.56$\pm$0.07 & $-$0.66$\pm$0.10 & $-$1.54$\pm$0.09 \\
CCM~53  & 0.51 & 8.66$\pm$0.09 & $-$0.63$\pm$0.14 & $-$1.63$\pm$0.13 \\
CCM~54  & 0.56 & 8.49$\pm$0.08 & $-$0.53$\pm$0.12 & $-$1.52$\pm$0.10 \\
CCM~55  & 0.51 & 8.60$\pm$0.08 & $-$0.49$\pm$0.12 & $-$1.44$\pm$0.10 \\
CCM~57  & 0.51 & 8.40$\pm$0.09 & $-$0.54$\pm$0.14 & $-$1.50$\pm$0.14 \\
CCM~57A & 0.72 & 8.58$\pm$0.14 & $-$0.65$\pm$0.17 & $-$1.61$\pm$0.24 \\
CCM~71A & 1.04 & 8.45$\pm$0.07 & $-$0.62$\pm$0.10 & $-$1.58$\pm$0.08 \\
CCM~72  & 0.51 & 8.71$\pm$0.10 & $-$0.63$\pm$0.15 & $-$1.56$\pm$0.13 \\
CCM~84A & 0.44 & 8.52$\pm$0.13 & $-$0.44$\pm$0.19 & $-$1.64$\pm$0.17 \\
P\,203  & 0.19 & 8.84$\pm$0.16 & $-$0.46$\pm$0.13 & $-$1.61$\pm$0.20 \\
\enddata
\end{deluxetable}

The gradient from equation (4) corresponds to $-0.02\pm 0.01$
dex\,kpc$^{-1}$.  The central O/H abundance ratio suggested by the
regression is about solar or slightly above solar, but we cannot
exclude a steepening of the relation at small galactocentric radii.
Clearly, data for additional \hii\/ regions are needed to better
constrain the abundance gradient in M51.  There is considerable
scatter at a given radial distance, amounting to $\sim$0.4 dex in O/H
around $R/R_0=0.5$. This is not atypical, since scatter of a similar
magnitude is observed in other well-studied spiral galaxies
(\citealt{kennicutt03}; \citealt{garnett97}). Finally, the radial
trend of both N/O and S/O are consistent with a constant value,
although a very shallow N/O gradient is suggested by
Fig.~\ref{radial}.

\begin{figure}[h]
\plotone{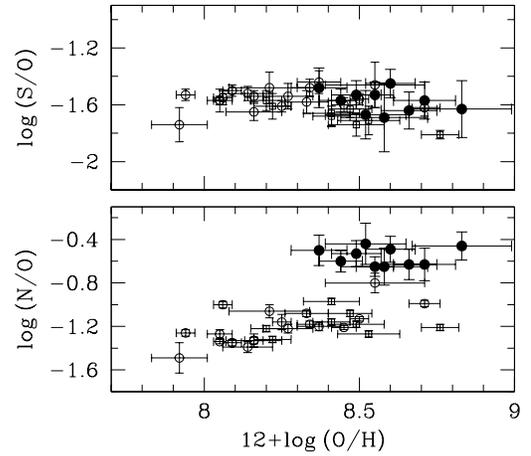}
\caption{{\em Top:} The relationship between the S/O abundance ratio
  and O/H. Symbols are the same as in Fig.~\ref{ion_s}. A trend of
  approximately constant S/O is seen to extend beyond the solar oxygen
  abundance. {\em Bottom:} The relationship between N/O and
  O/H. Symbols as in Fig.~\ref{ion_s}.}
\label{sulfur}
\end{figure}

Plots of the abundance ratios S/O and N/O as a function of O/H are
shown in Fig.~\ref{sulfur}. The S/O ratio (upper panel) for the M51
\hii\/ regions is consistent with the approximately constant value
[log\,(S/O)\,=\,$-1.6$] observed in other spiral galaxies, without any
evidence for a decrease as the oxygen abundance increases. The results
of \citet{diaz91} had provided some suggestion that such a decrease
might occur (e.g.~\citealt{garnett03}). Based on our results, we
therefore find no need to invoke changes in the massive star initial
mass function or in the nucleosynthesis at high metallicity (at least
up to approximately the solar value).

The nitrogen abundance, shown in relation to O/H in the bottom panel
of Fig.~\ref{sulfur}, has an average value log\,(N/O)\,$\simeq -0.56$,
higher than the value measured in \hii\/ regions of comparable oxygen
abundance in galaxies like M101 and NGC~2403. This apparent nitrogen
overabundance seems consistent with the known spread of N/O at a given
O/H for spiral galaxies, related to differences in star formation
histories, with earlier Hubble types (M51 is classified as Sbc) having
on average higher N/O ratios than later types (\citealt{henry00};
\citealt{pilyugin03}).


\section{Empirical abundance indicators}

The results of a quarter century of research on nebular abundances
have proved very useful to obtain estimates of abundance gradients in
spiral galaxies and of oxygen abundances of star forming regions in
the distant universe. Indirect methods (as opposed to the direct
methods based on the determination of electron temperatures) using
ratios of strong lines have been developed by several authors, the
most widely used being arguably the $R_{23}$ empirical abundance
indicator of \citet{pagel79}. Some evidence has been mounting,
however, against the accuracy of such methods in the abundance range
typical of the central regions of spiral galaxies (half solar oxygen
abundance and above). This has been discussed at length in our work on
M101 (\citealt{kennicutt03}), and we will not repeat the discussion
here. In that paper we also presented the main issues which could
complicate the analysis based on the collisionally excited lines,
namely the presence of temperature fluctuations and, at high
metallicity, temperature gradients. Their effect is similar, in that
the measured electron temperature is weighted towards the higher
temperature zones, leading to a possible underestimate of the real
nebular abundances. The reader is referred to \citet{kennicutt03} for
further details and for a justification for the use of the
collisionally excited lines.  Admittedly, the issue is still open.  In
addition, the temperature in the \nii\/ zone could be overestimated
from neglecting the recombination component of the \lin5755 line (see
\citealt{liu00}). The lack of the line flux from \niii\/ at 57 $\mu$m
for our M51 sample does not allow us to estimate the contribution of
this recombination directly from observations. However, this
contribution is likely to be small, given the small amount of N$^{++}$
present in low-excitation nebulae. The good agreement between the
observed T\nii-T\siii\/ relationship and the predicted one from
\citet{garnett92} also suggests that such a recombination component
can be neglected to first order.

The detection of auroral lines in objects around the solar O/H ratio
is prompting a revision of the abundances based on indirect methods,
since relatively low abundances are found for objects that were once
thought to be very metal-rich (i.e.~well above the solar value:
\citealt{castellanos02}, \citealt{kennicutt03}). This is also the case
for the M51 \hii\/ regions analyzed in this work, and a brief
assessment of a few representative empirical abundance indicators
seems to be warranted at this point.

\begin{figure}[h]
\plotone{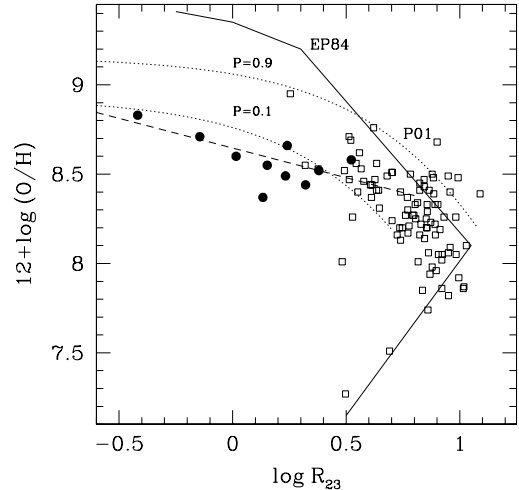}
\caption{Comparison between O/H abundances obtained from measured
electron temperatures (dots) and from the semi-empirical abundance
indicator $R_{23}$ (full and dotted lines). Two different $R_{23}$
calibrations are used: Edmunds \& Pagel (1984,\,=\,EP, full line) and
Pilyugin (2001,\,=\,P01, dotted lines). Two curves are plotted for the
latter, corresponding to two different values of the excitation
parameter, P\,=\,0.1 and P\,=\,0.9.  The plotted data are from the
current work (full dots) and from published studies:
\citet{kennicutt03}, \citet{garnett97}, \citet{vanzee98},
\citet{deharveng00}, \citet{diaz00} and \citet{castellanos02} (open
squares). The lowest excitation object, at the far left, is P\,203,
while the most metal-rich one is CDT1 in NGC~1232
(\citealt{castellanos02}).  The dashed line is a fit to the M51
points, except CCM~84A: $12\,+\log{\rm (O/H)}=8.64 - 0.33~\log {\rm
R}_{23}$.}
\label{r23}
\end{figure}

It is obvious that the abundances obtained in M51 in the present work
have an important impact on most empirical calibrations. This is
exemplified in Fig.~\ref{r23}, where we show the comparison of our
direct oxygen abundances with those obtained by means of $R_{23}$,
adopting two different calibrations: the one by \citet{edmunds84} and
the one by \citet{pilyugin01} (see \citealt{garnett04} for a
comparison involving two other calibrations). The sample of objects
shown includes, besides the M51 \hii\/ regions studied here, a number
of nebulae extracted from the literature (\citealt{kennicutt03};
\citealt{garnett97}; \citealt{vanzee98}; \citealt{deharveng00};
\citealt{diaz00}; \citealt{diaz00b}; \citealt{castellanos02}).  We
note that for the high-metallicity object CDT1 in NGC~1232 studied by
\citet{castellanos02} we derive, using their line fluxes, an O/H value
smaller by 0.13 dex than their published abundance
12\,+\,log\,(O/H)\,=\,8.95.  Our results on the M51 \hii\/ regions
clearly demonstrate the failure of the existing $R_{23}$ calibrations
at low excitation (log\,$R_{23}<0.5$), since they overestimate the
abundances by 50--400 percent. This discrepancy corresponds to
electron temperatures smaller by approximately 10--15 percent compared
with our measured values (about 3$\sigma$ to 4$\sigma$). In the
particular case of CCM~72 and CCM~10, the \oii\/ temperatures modeled
by \citet{diaz91} (4800~K and 6000~K, respectively) lie 4$\sigma$ and
6$\sigma$ below our measurements.

\begin{figure}[h]
\plotone{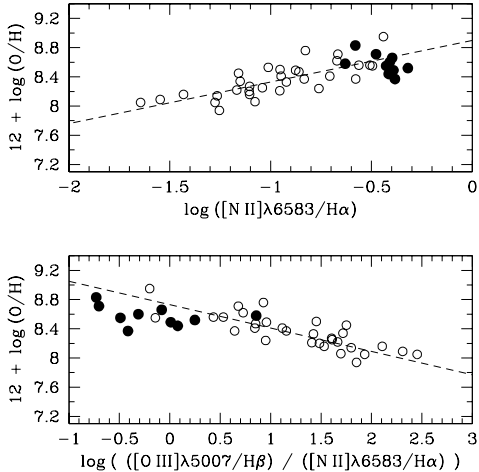}
\caption{{\em Top:} O/H abundances obtained with the direct method are
 plotted against the line ratio \nii\lin 6583/H$\alpha$. Data are from
 this work (full dots) and from \citet{kennicutt03}, \citet{garnett97}
 and \citet{castellanos02} (open circles). The dashed line is the
 regression taken from \citet{pettini04}, obtained from a fit to a
 larger sample of \hii\/ regions. {\em Bottom:} Same as above, for the
 (\oiii\lin 5007/H$\beta$)/(\nii\lin 6583/H$\alpha$) index.}
\label{pettini}
\end{figure}

\begin{figure}[h]
\plotone{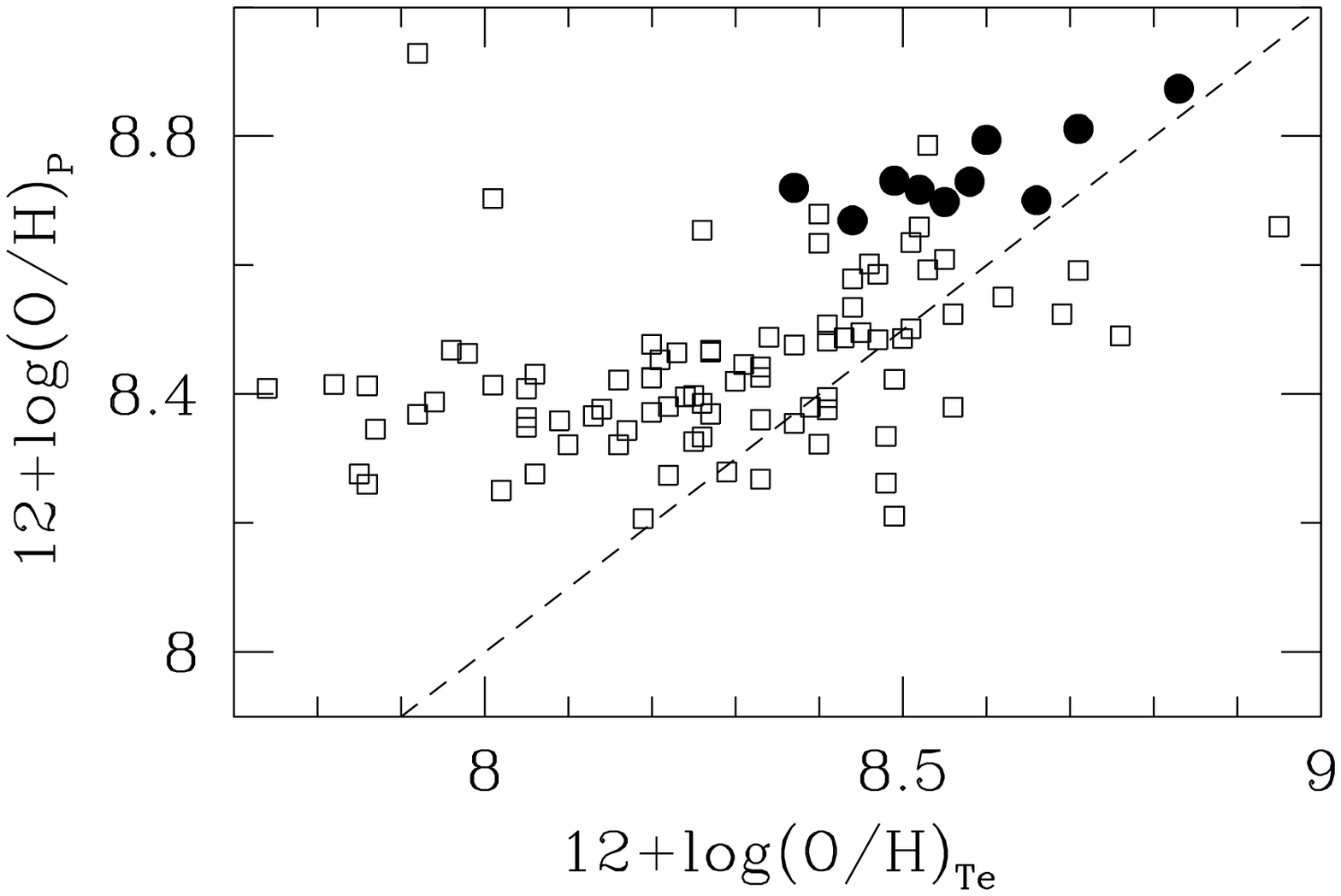}
\caption{Comparison between direct O/H abundances and those obtained
  through the P-method of \citet{pilyugin01}. The data are from the
  current work (full dots) and from \citet{kennicutt03},
  \citet{garnett97}, \citet{vanzee98}, \citet{deharveng00},
  \citet{diaz00} and \citet{castellanos02} (open squares). Only
  objects which are predicted by the P-method to have
  12\,+\,log\,(O/H)$>$8.2 are plotted.  }
\label{pily}
\end{figure}

This result has fundamental consequences for the studies of abundance
gradients in spiral galaxies based on this empirical indicator
(e.g.~\citealt{zaritsky94}, \citealt{vila92}), and the actual
gradients in the central regions will in general be much flatter than
reported in these works.  The discrepancy is also not peculiar to
$R_{23}$. As an example, \citet{pettini04} have considered an
empirical abundance calibration of two line indices involving the
nitrogen lines: \nii\lin 6583\,/\,H$\alpha$ (see also
\citealt{denicolo02}), and (\oiii\lin 5007/H$\beta$)\,/\,(\nii\lin
6583/H$\alpha$). We show in Fig.~\ref{pettini} the position occupied
in the corresponding diagrams by the M51 \hii\/ regions. A decrease in
the slope at low excitation is noticed with respect to the
\citet{pettini04} calibrations (shown by the dashed lines), indicating
a flattening around 12+log\,(O/H)\,=\,8.6, which however also reflects
the large nitrogen abundance found in M51 from our analysis.
Fig.~\ref{pily} shows instead how the O/H abundances derived through
the P-method (\citealt{pilyugin01}) compare with direct abundances. In
this plot, only objects which are predicted by the P-method to have
12\,+\,log\,(O/H)$>$8.2 (the quoted range of validity for this method
in the high-abundance regime) are plotted.  This indicator seems to be
afflicted by severe difficulties in a wide abundance range.  The
general conclusion we can draw is that great caution is necessary when
using strong-line empirical abundance indicators for low-excitation
\hii\/ regions. Erroneous results can be obtained in the inner disks
of spirals up to a factor of a few.

How low can the excitation in M51 \hii\/ regions be? We report in
Table~6 the line fluxes, normalized to H$\beta=100$, measured for the
\hii\/ regions belonging to the central zone of the galaxy ($R_0=0.12
- 0.36\; R_0$) besides P\,203. No auroral line was detected in these
objects, therefore not permitting a direct abundance
determination. The total \oiii\lin\lin4959,5007 flux is just a few
percent of the H$\beta$ line flux, and the corresponding value of
$\log R_{23}$ is in the range $-0.42$ to $-0.67$.  Because of the
rather flat relationship between O/H and $R_{23}$ at low excitation
suggested by the M51 points in Fig.~\ref{r23}, these very low $R_{23}$
values might still correspond to metallicities not much higher than
solar [12\,+\,log\,(O/H)\,$\approx$\,8.8--8.9], however any
extrapolation to determine an oxygen abundance is at present still
very uncertain. The dashed line in Fig.~\ref{r23} shows such an
extrapolation, determined from the M51 \hii\/ regions. The highest
oxygen abundance in our sample, corresponding to $\log R_{23}=-0.66$
(the value for CCM~6A, CCM~37A and CCM~107 within $\pm0.01$), would be
12\,+\,log\,(O/H)\,$\approx$\,8.86 (1.5 times the solar O/H),
comparable to our measured oxygen abundance for P\,203.

We have estimated upper limits for the \nii\/ electron temperature and
the corresponding lower limits for the oxygen abundance from the
non-detection of the \nii\lin 5755 line in the spectra of the central
objects. The results, summarized in Table~6, are not very compelling,
and they are still consistent with abundances in the central region of
M51 around the value found for P\,203.

\begin{figure}[h]
\plotone{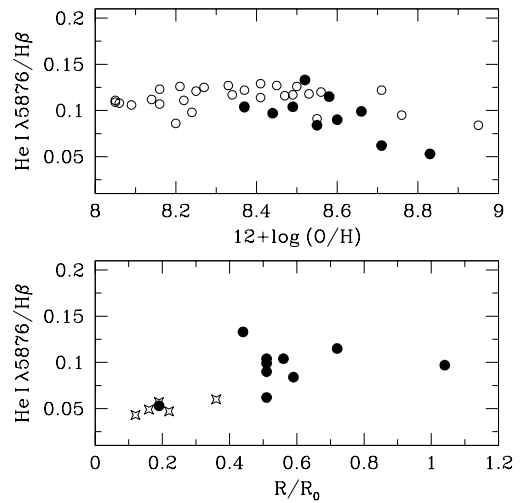}
\caption{{\em Top:} A trend of decreasing He\,{\sc i}\lin
5876/H$\beta$ line ratio with abundance above
12\,+\,log\,(O/H)\,$\approx$\,8.5 is suggested here. The \lin 5876
line has been corrected for an underlying absorption equivalent width
of 0.8~\AA. Objects plotted are from M51 (filled dots) and M101 (open
circles). {\em Bottom:} Variation of He\,{\sc i}\lin 5876/H$\beta$
with galactocentric distance for M51 \hii\/ regions. The filled dots
are objects with a direct abundance determination, the open stars are
objects of low excitation found in the central region of the galaxy.}
\label{he}
\end{figure}

We finally point out that the He\,{\sc i}\lin 5876/H$\beta$ line ratio
correlates with both O/H (for the objects in Table~1) and
galactocentric distance (objects in Tables~1 and 6), as shown in
Fig.~\ref{he}. Together with the reduced excitation measured from the
\oiii/H$\beta$ ratio mentioned above, Fig.~\ref{he} shows how the
ionizing field becomes softer towards higher metallicities, above a
certain abundance threshold (the He\,{\sc i}\lin 5876/H$\beta$ ratio
decreases below its saturation level, at which He is completely singly
ionized, for 12\,+\,log\,(O/H)\,$\geq$\,8.5, top panel) and towards
the galactic central regions (lower panel).  The observed increase of
the neutral helium fraction towards the metal-rich central region of
M51 can be viewed as the result of the decrease in the equivalent
effective temperature of the ionizing radiation field with increasing
metallicity (see also \citealt{bresolin02}).

\begin{deluxetable*}{lcccccccccc}
\tabletypesize{\scriptsize}
\tablecolumns{11}
\tablewidth{0pt}
\tablenum{6}
\tablecaption{Line fluxes of low-excitation objects}
\tablehead{
\colhead{ID}            &
\colhead{$R/R_0$}            &
\colhead{C(H$\beta$)}    &
\colhead{EW(H$\beta$)}    &
\colhead{[O\,{\sc ii}]}    & 
\colhead{[O\,{\sc iii}]}    & 
\colhead{He~{\sc i}}       &
\colhead{[N\,{\sc ii}]}    & 
\colhead{[S\,{\sc ii}]}   &
\colhead{T\nii}         &
\colhead{12\,+\,log\,(O/H)}    \\
\colhead{}        &
\colhead{}        &
\colhead{}        &
\colhead{(\AA)}        &
\colhead{3727}    & 
\colhead{4959+5007}    & 
\colhead{5876}    &
\colhead{6548+6584}    & 
\colhead{6717+6731}    &
\colhead{(upper limit)}  &
\colhead{(lower limit)} }
\startdata
\vspace{-2mm}
\\
CCM~6A  & 0.12 & 0.10 & 55 & 17$\pm$1 & 4.4$\pm$0.3 & 3.9$\pm$0.3 & 80$\pm$4 & 26$\pm$1 & 5800~K & 8.40 \\
CCM~37A  & 0.16 & 0.09 & 12 & 17$\pm$1 & 4.8$\pm$0.4 & 3.0$\pm$0.3 & 54$\pm$3 & 20$\pm$1 & 6200~K & 8.20 \\
CCM~45  & 0.36 & 0.17 & 54 & 34$\pm$2 & 4.2$\pm$0.4 & 5.5$\pm$0.5 & 116$\pm$6 & 50$\pm$3 & 5900~K & 8.52 \\
CCM~56  & 0.22 & 0.23 & 50 & 28$\pm$2 & 8.5$\pm$0.9 & 4.5$\pm$0.5 & 121$\pm$7 & 50$\pm$3 & 5900~K & 8.58 \\
CCM~107  & 0.19 & 0.53 & 48 & 20$\pm$1 & 2.6$\pm$0.2 & 5.2$\pm$0.3 & 98$\pm$5 & 30$\pm$2 & 5300~K & 8.86 \\
\enddata
\end{deluxetable*}


\section{Wolf-Rayet stars}
\begin{figure}[ht]
\plotone{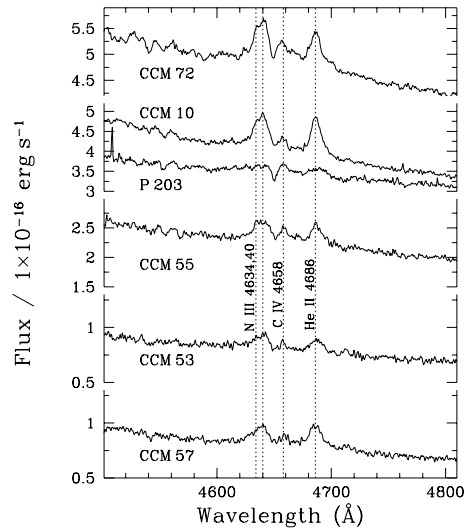}
\caption{The WR stellar lines around 4660~\AA\/ have been detected in
  these six \hii\/ regions out of the ten analyzed. Feature
  identification is provided by the vertical dotted lines.}
\label{wr}
\end{figure}

The main spectroscopic signature of WR stars (the blue bump at
4660~\AA) has been detected in six of the \hii\/ regions examined in
this work. As shown in Fig.~\ref{wr} we can clearly distinguish the
stellar N\,{\sc iii}\lin 4634--41 and He\,{\sc ii}\lin 4686 lines in
all cases, together with C\,{\sc iv}\lin 4658.  Given the absence of
N\,{\sc v}\lin 4603--20, the morphology of the blue bump is
characteristic of late WR subtypes (WNL). In addition, the C\,{\sc
iii}\lin 5696 line from WC subtype stars has been clearly detected in
the two most metal-rich \hii\/ regions, CCM~72 and P\,203.  WR
features have been detected in other M51 \hii\/ regions not analyzed
in this paper. In CCM~13, 24, 41, 84A and 91 only the blue bump is
present in our spectra. Both the 4660~\AA\/ and the 5696~\AA\/
features are seen in CCM~37A, 84 and 107, while only the 5696~\AA\/
line is visible in CCM~6A and 71 (also with 5810~\AA).  A marginal
detection of He\,{\sc ii}\lin 4686 has been made in a few additional
objects, as well. Note that all the nebulae (except CCM~71) where the
C\,{\sc iii}\lin 5696 line has been detected are located at small
galactocentric radii, or have a large O/H abundance (CCM~72). This
agrees with the known trend of increasing fraction of WC relative to
WN stars with abundance (\citealt{massey03}).

The number of WNL stars can be estimated from the total He\,{\sc
ii}\lin 4686 line luminosity, assuming a fixed stellar luminosity of
$1.6\times 10^{36}$ erg\,s$^{-1}$ (\citealt{schaerer98}). At the
adopted distance to M51 of 8.4 Mpc the two objects with the highest
He\,{\sc ii} luminosity (CCM~10 and CCM~72) are then found to contain
five-six WNL stars each. The luminosity of the remaining objects is
consistent with just one or two WNL stars being present.  These
numbers are not particularly remarkable when compared, for example,
with the very large population of WR stars discovered by
\citet{crowther04} in \hii\/ regions of another supposedly metal-rich
galaxy, M83.  Reporting the detection of WR features in solar-like
abundance environments, however, is still of considerable interest,
because of the trends one can establish between WR star population
properties (e.g.~their subtype distribution) and metallicity. It is also
relevant to point out that WR features are found in very cool nebulae,
such as the metal-rich CCM~72 and P\,203, in support of the idea that
the presence of these evolved massive stars does not significantly
affect the ionizing properties of the embedded star clusters
(\citealt{bresolin02}).

\section{Gas Fractions, Effective Yields, and Chemical Evolution}

A comparison of observed abundances and gas fractions in galaxies
provides interesting information on the effect of gas flows on
galaxy evolution. Variations of the effective yield, defined as

\begin{equation}
y_{\!\mbox{\scriptsize\rm eff}} = {Z \over {\ln (\mu^{-1})} },
\end{equation}

\noindent
where $Z$ is the metallicity and $\mu$ is the gas mass fraction, are
sensitive to gas inflow and outflow (\citealt{edmunds90,
koppen99}). \citet{garnett02} compared effective yields, derived from
global abundance, stellar mass, and gas mass properties, in a sample
of 50 nearby spiral and irregular galaxies, and found that \yeff\/
varies systematically with galaxy rotation speed (and thus presumably
total mass). The adopted interpretation was the low-mass irregular
galaxies, with smaller effective yields than those derived for massive
spirals, were losing a large fraction of their metals to the
intergalactic medium (IGM).

Based on our new abundances for M51, we can re-evaluate its effective
yield. \citet{garnett02} used the measured abundance at one effective
disk radius as a measure of the average metallicity of a spiral
disk. From his Table 4, the adopted average 12 + log\,(O/H) was 9.12
for M51. Our new value for 12 + log\,(O/H) at
$R_{\!\mbox{\scriptsize eff}}$ = 0.6\,$R_0$ is 8.55, a factor of
3.7 smaller. Thus the new value of \yeff\/ we would derive is smaller
by the same amount. \citet{garnett02} derived \yeff\/ = 0.012 for M51,
so we would obtain \yeff\/ = 0.0032 from our new data. This lower
value is comparable to the values derived for smaller irregular
galaxies such as IC~10 and NGC~6822 in Garnett's sample. Note also
that this value of \yeff\/ is much closer to that expected from
stellar nucleosynthesis, and to the value obtained from estimates of
the mean metallicity and gas fraction in the solar neighborhood. If
other spirals show similar reductions in \yeff\/ when more
temperature-based abundance measurements become available, it may be
necessary to re-examine the conclusions of \citet{garnett02} regarding
what galaxies may experience significant loss of metals.

It is also interesting and important to look at the variation of
\yeff\/ across a disk galaxy, to see if gas flows may play a role in
the chemical evolution within the disk. A comparison between gas
fraction and abundance across the disk has been done for only a few
spirals to date.

We obtained the radial distribution of gas surface densities in M51
from the single-dish CO mapping of \citet{kuno95}, measured with a
16$\arcsec$ beam, and from the \hi\/ 21-cm map of
\citet{tilanus91}. \citet{young95} also measured the radial
distribution of CO, but their much larger beam size means that the
mass surface density is averaged over a much larger area. We choose
not to use measurements based on interferometers
(e.g.~\citealt{aalto99}), as they can miss significant amounts of
extended emission due to missing short antenna baselines; we note that
the measurements of \citet{aalto99} show a very sharply peaked CO flux
distribution in rough agreement with \citet{kuno95}. The derived
molecular gas surface densities depend critically on the conversion
from CO intensity to H$_2$ column density. We derived molecular
surface densities under two assumptions for the I(CO)-N(H$_2$)
conversion factor: (1) the `standard' factor 3.0$\times$10$^{20}$
cm$^{-2}$ (K km s$^{-1}$)$^{-1}$ (\citealt{wilson95}), which includes
a contribution for helium; (2) the smaller factor 1.3$\times$10$^{20}$
cm$^{-2}$ (K km s$^{-1}$)$^{-1}$ used by \citet{kuno95} (adjusted here
to include helium).  Whichever value is used, the mass of molecular
gas in M51 is quite large, and the mass surface density of H$_2$
exceeds that of atomic gas over much of the disk. The CO measurements
do not go beyond 180$\arcsec$ from the nucleus, so we have smoothly
extrapolated the H$_2$ surface densities to larger radii based on an
approximate exponential fit to the inner disk profile. Gas surface
density distributions for a subset of our assumptions are shown in
Figure~12.

\begin{figure}[ht]
\plotone{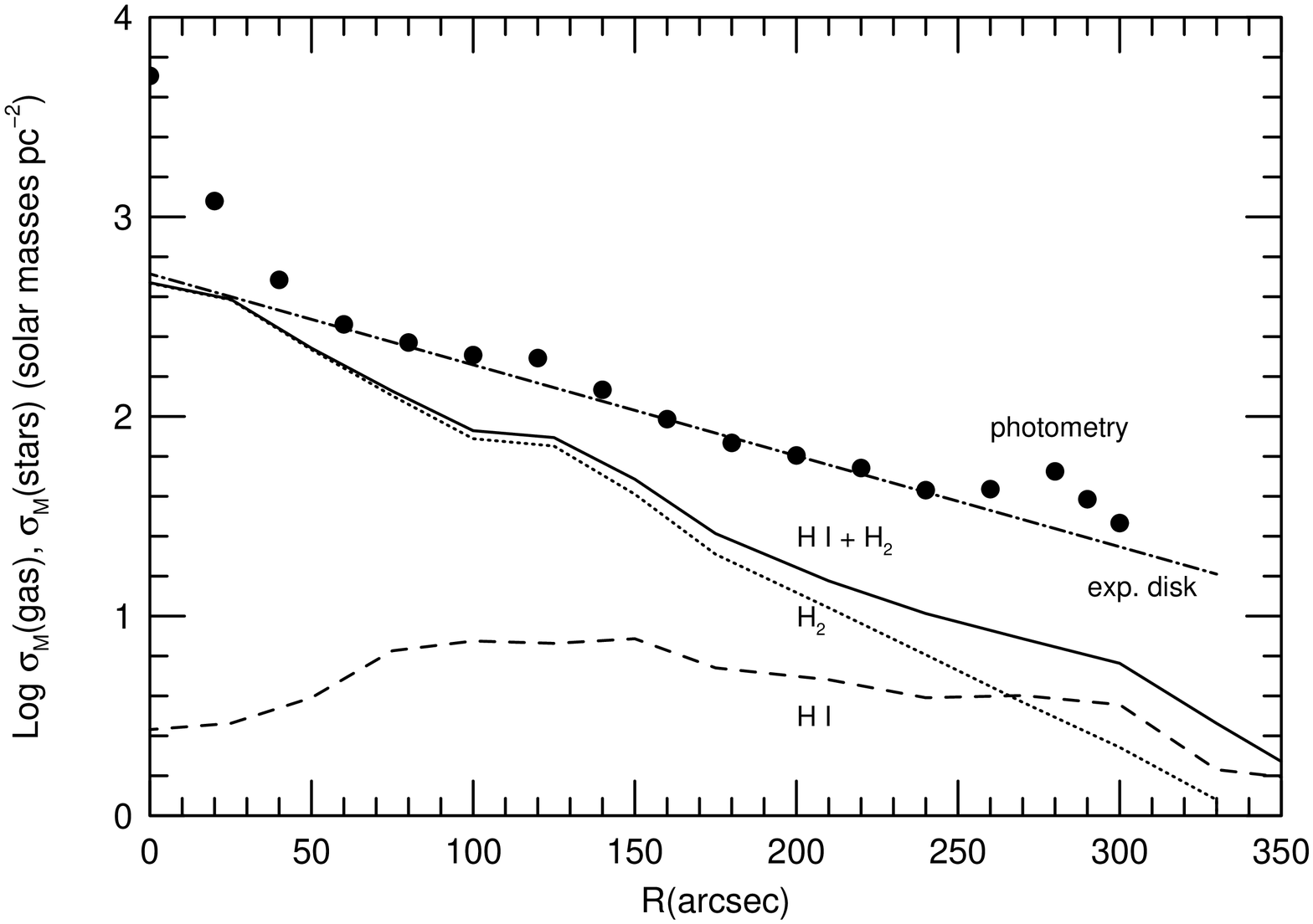} \figcaption{Gas and stellar mass surface density
distributions in M51. The dashed line shows the \hi\/ surface density
(adjusted for the presence of helium) from the data of
\citet{tilanus91}.  The dotted line shows the molecular gas profile
obtained from the CO measurements of \citet{kuno95}, assuming a
CO-H$_2$ conversion factor of 3$\times$10$^{20}$ cm$^{-2}$ (K km
s$^{-1}$)$^{-1}$. The dot-dash line shows the mass profile for the
stellar exponential disk determined as discussed in the text. The
filled circles show stellar mass densities derived directly from the
surface brightness data of \citet{kuchinski00}.}
\label{don1}
\end{figure}

The mass surface density for stars in M51 was derived from the BVRI
surface photometry of \citet{kuchinski00}. We converted the I-band
surface brightnesses to mass surface densities by estimating the
mass-to-light ratio (M/L) from the relations between M/L and color
derived from stellar population models in \citet{bell01}.  From the
Kuchinski et al. surface brightnesses, we estimated a central I-band
disk surface brightness of 19.05 mag arcsec$^{-2}$ and a disk scale
length of 114$\arcsec$; we adopted a linear color profile with radius
ranging from B--R = 1.2 at the center to B--R = 0.9 at 300$\arcsec$.
The resulting stellar mass profile for the disk in M51 is shown in
Figure~12 as the dot-dash line, while the mass densities estimated
directly from the photometry are shown as the filled circles. Note
that the photometry shows evidence for a central spheroidal component.

Inspection of Figure~12 shows one curious feature: the gas surface
density profile is steeper than the stellar disk profile. This implies
that the gas fraction decreases radially outward. We illustrate this
in Figure~13(a), which shows the radial variation in three different
cases: (1) I(CO)-N(H$_2$) = 3.0$\times$10$^{20}$ (solid line); (2)
I(CO)-N(H$_2$) = 1.3$\times$10$^{20}$ (dashed line); and (3) a case
similar to (1), but in which the stellar surface brightnesses have
been adjusted upward for an extinction A(I) = 0.5 mag (dotted
curve). In all three cases, the derived gas fraction declines by a
factor 3-4 radially outward. This behavior contrasts strongly with
that seen in other spiral galaxies, where the gas fraction increases
outward (M81 - \citealt{garnett87}; NGC 4254 - \citealt{henry94}; NGC
2403 and M33 - \citealt{garnett97}).

We have also looked at what happens if we simply use the observed
surface brightness profile to derive the stellar mass densities,
rather than an underlying exponential disk. The resulting gas profiles
are shown in Figure~13(b), for our two assumptions of the
I(CO)-N(H$_2$) conversion. While the resulting gas fractions for the
inner disk are certainly smaller, due to the high central surface
brightness, we still see a declining gas fraction with radius for R
$>$ 50$\arcsec$.

\begin{figure}
\epsscale{0.9}
\plotone{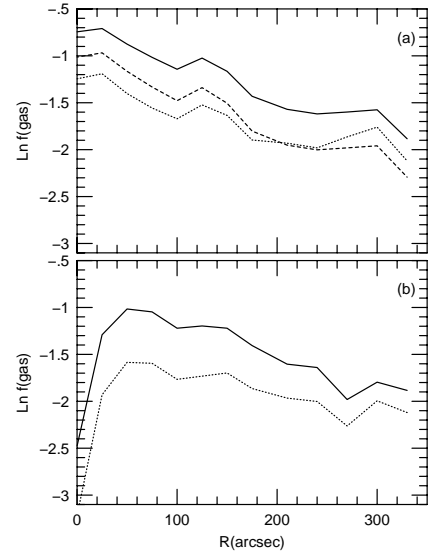} \figcaption{(a) The radial variation of the gas
mass fraction for three different assumptions. {\em Solid line:}
molecular gas masses estimated using I(CO)-N(H$_2$) conversion of
3.0$\times$10$^{20}$; {\em dashed line:} molecular gas mass densities
estimated using I(CO)-N(H$_2$) conversion of 1.3$\times$10$^{20}$;
{\it dotted line:} same gas densities as for solid line, but with
stellar surface brightnesses adjusted for extinction A(I) = 0.5 mag.
(b) Radial variation of the gas fraction, where the stellar mass
densities were derived directly from the I-band photometry, rather
than the underlying exponential disk.  {\em Solid line:} molecular gas
densities obtained using I(CO)-N(H$_2$) = 3.0$\times$10$^{20}$; {\em
dotted line:} molecular gas densities obtained using I(CO)-N(H$_2$) =
1.3$\times$10$^{20}$.}
\label{don2}
\epsscale{1.0}
\end{figure}

We note that a radial variation of a factor three in the CO-H$_2$
conversion is not sufficient to mitigate the decline in gas fraction.
A factor six change in the conversion factor would be needed to
flatten the gas fraction profile; such a variation is not at all
predicted even by the most extreme estimates of metallicity variation
in the conversion factor (\citealt{israel97}). A radial decline of
about 1.5 mag in A(I) from the center to the outer disk would also be
enough to flatten the gas fraction profile. \citet{scoville01} found a
rough radial variation in A(V) for \hii\ regions in the disk of M51,
from A(V) $\approx$ 4 in the inner disk to A(V) $\approx$ 2 at R =
300$\arcsec$; however, it is not clear how to relate the extinction
for \hii\ regions, which are associated with gas and dust, to the
extinction for the stellar disk. We also note that A(I) = 0.5
corresponds to E(B--V) = 0.3; with an observed B--V of 0.6, such a
correction for reddening would make M51 a very blue galaxy unless the
reddening is very gray.

A radially decreasing gas fraction combined with a radially decreasing
metallicity implies an effective yield that increases with
metallicity.  Such a variation was inferred by \citet{vila92}, but
with older abundance data derived mainly from strong emission line
calibrations. On the other hand, galaxies with abundances based on
electron temperatures show little evidence for variations in effective
yields (\citealt{garnett97}).

The peculiar gas fraction profile in M51 may also point to dynamical
influences. M51 is an interacting system. Fly-by interactions of this
type can lead to bar formation, which induces radial flows of gas
toward the center of the galaxy. \citet{pierce86}, \citet{kohno96},
and \citet{aalto99} have presented photometric and kinematic evidence
for a bar in the central regions of NGC 5194, while infrared imaging
reveals that NGC 5195 has an obvious bar (e.g., the I-band images of
\citealt{kuchinski00}). It is possible that the interaction and bar
have driven gas into the central regions of M51, leading to the very
steep gas profile we see. This would complicate the interpretation of
the chemical evolution of this system, and requires dynamical and
photometric modeling that is beyond the scope of this work.


\section{Conclusions}

We have analyzed a sample of ten \hii\/ regions in the spiral galaxy
M51 where at least one of the auroral lines \nii\lin 5755 and
\siii\lin 6312 could be measured, and we discussed the resulting
oxygen, sulfur and nitrogen abundances. The single, most remarkable
result obtained is that in M51 the direct O/H abundances are
considerably below the value obtained either from photoionization
models (\citealt{diaz91}) or from empirical abundance indicators. The
calibration of the latter depended so far, in the low-excitation range
defined  mostly by \hii\/ regions in M51, on the abundances
obtained from the models, therefore the two are not independent. We
have not attempted to generate improved nebular models, for example
including more recent stellar atmospheres for the treatment of stellar
ionizing fluxes. This aspect needs to be explored in the future.

The result concerning O/H has important implications for the
calibration of empirical abundance indicators, and in general for the
determination of abundance gradients in spiral galaxies. Our result,
combined with previous ones obtained for smaller samples of metal-rich
\hii\/ regions by \citet{castellanos02}, \citet{diaz00b} and
\citet{kennicutt03}, points to shallower gradients. \hii\/ regions
once believed to have O/H abundances equal to 2--3 times solar are
found by the direct method to rarely exceed the solar value, and if so
by a moderate amount, up to 50 percent.

Our analysis has also established that in our M51 sample the abundance
ratio S/O is similar to the value found at lower O/H abundance in
other spiral galaxies. We find no evidence for a decrease in S/O with
increasing O/H, which would occur in case of a differing initial mass
function and/or nucleosynthesis for massive stars. The nitrogen
abundance has a mean value log\,(N/O)\,$\simeq$\,$-0.6$, larger than
in later spirals like M101, but this can still be interpreted with the
known spread of N/O at a given oxygen abundance.

The lower oxygen abundance we measure in M51 with respect to previous
determinations has also allowed us to revise the effective yield of
this galaxy. From a rather peculiar gas fraction radial profile we
infer an effective yield increasing with metallicity.

The number of galaxies with reliable {\em direct} measurements of
nebular chemical abundances extending over their whole disks,
including the metal-rich central zones, is still very limited. These
measurements are necessary for the correct determination of the
chemical compositions in galactic disks and their interpretation in
terms of chemical evolution models.  The current work on M51
demonstrates that large samples of metal-rich \hii\/ regions can be
efficiently studied in spiral galaxies with current telescopes and
instrumentation.

\acknowledgments We thank T. Rector for the Kitt Peak 0.9m H$\alpha$
image of M51 used for Fig.~1. RCK acknowledges NSF grant AST-0307386
and NASA grant NAG5-8426. DRG acknowledges NSF grant AST-0203905 and
NASA grant NAG5-7734.


\end{document}